\documentclass[superscriptaddress,twocolumn,prl]{revtex4}
\usepackage[utf8]{inputenc}
\usepackage[english]{babel}
\usepackage[T1]{fontenc}
\usepackage{amsmath}
\usepackage{amsfonts}
\usepackage{amssymb}
\usepackage{xcolor}
\usepackage[colorlinks=true,linkcolor=blue,filecolor=blue, urlcolor=blue,citecolor=blue]{hyperref}
\usepackage{graphicx}
\usepackage{float}
\usepackage{color}
\usepackage{soul}
\usepackage{siunitx}
\usepackage{physics}
\usepackage{lineno}
\usepackage{subfiles}

\begin{document}

\title{Photoexcited Hole States at the SrTiO$_3$(001) Surface Imaged with Noncontact AFM}

\author{Igor Sokolović}
\affiliation{Institute of Applied Physics, TU Wien, 1040 Vienna, Austria}

\author{Florian Ellinger}
\affiliation{University of Vienna, Faculty of Physics and Center for Computational Materials Science, Vienna, Austria}

\author{Aji Alexander}
\affiliation{Department of Surface and Plasma Science, Faculty of Mathematics and Physics, Charles University, 180 00 Prague 8, Czech Republic}

\author{Dominik Wrana}
\affiliation{Department of Surface and Plasma Science, Faculty of Mathematics and Physics, Charles University, 180 00 Prague 8, Czech Republic}
\affiliation{Marian Smoluchowski Institute of Physics, Jagiellonian University, Krakow 30-348, Poland}

\author{Lloren\c{c} Albons}
\affiliation{Department of Surface and Plasma Science, Faculty of Mathematics and Physics, Charles University, 180 00 Prague 8, Czech Republic}

\author{Sreehari Sreekumar}
\affiliation{Department of Surface and Plasma Science, Faculty of Mathematics and Physics, Charles University, 180 00 Prague 8, Czech Republic}

\author{Michael Schmid}
\affiliation{Institute of Applied Physics, TU Wien, 1040 Vienna, Austria}

\author{Ulrike Diebold}
\affiliation{Institute of Applied Physics, TU Wien, 1040 Vienna, Austria}

\author{Michele Reticcioli}
\affiliation{Institute of Applied Physics, TU Wien, 1040 Vienna, Austria}
\affiliation{University of Vienna, Faculty of Physics and Center for Computational Materials Science, Vienna, Austria}
\affiliation{CNR-SPIN, c/o Department of Physical and Chemical Sciences, University of L’Aquila, Via Vetoio, L’Aquila I-67100, Italy}

\author{Cesare Franchini}
\email[Corresponding author: ]{cesare.franchini@univie.ac.at}
\affiliation{University of Vienna, Faculty of Physics and Center for Computational Materials Science, Vienna, Austria}
\affiliation{Dipartimento di Fisica e Astronomia, Universit\`a di Bologna, 40127 Bologna, Italy}

\author{Martin Setvin}
\email[Corresponding author: ]{martin.setvin@matfyz.cuni.cz}
\affiliation{Department of Surface and Plasma Science, Faculty of Mathematics and Physics, Charles University, 180 00 Prague 8, Czech Republic}
\affiliation{Institute of Applied Physics, TU Wien, 1040 Vienna, Austria}

\begin{abstract} 
	
The behaviour of excess charges in ionic lattices, such as the formation of polarons and charge trapping at defect sites, influences the physical and chemical properties of materials and translates into applications in electronics, optics, photovoltaics, and catalysis. Here we show that the bulk-terminated SrTiO$_3$(001) surface accumulates photoexcited charges and keeps the associated photovoltage for many days at cryogenic temperatures. A combination of scanning tunneling microscopy, atomic force microscopy (STM/AFM) and Kelvin probe force microscopy (KPFM) was used to measure this photovoltage and to localize the photoexcited charges with atomic precision down to the single-quasiparticle limit. Density functional theory (DFT) shows that holes favor localization at oxygen $2p$ orbitals adjacent to Sr vacancies, creating long-lived trapped states. The methodology presented here provides guidelines for imaging of charges trapped in the crystal lattice using noncontact AFM. 

\end{abstract}

\maketitle

\begin{figure*}[t]
	\begin{center}
		\includegraphics[width=2.0\columnwidth,clip=true]{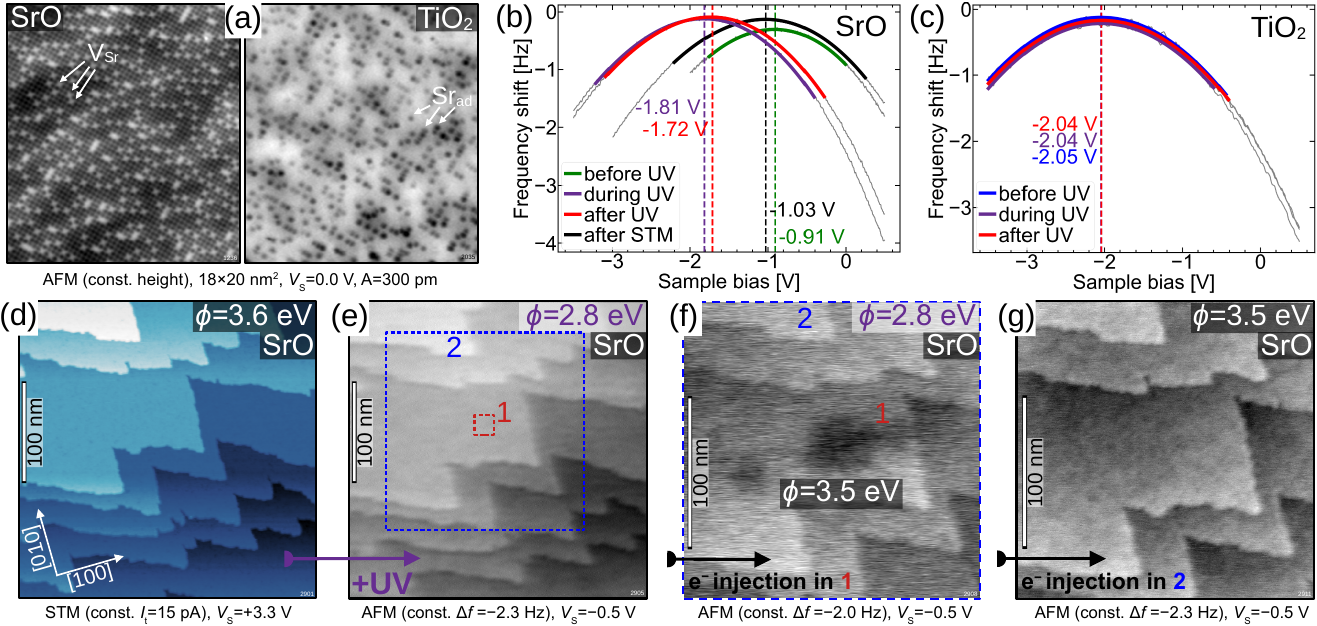}
	\end{center}
	\caption{			
		\textbf{Changes in work function of the SrO termination during UV irradiation.}
		a) Atomically resolved images of cleaved SrTiO$_3$, showing bulk-like ($1\times 1$) termination. Dark colors (low frequencies) indicate stronger attractive tip-surface interaction. The SrO termination (left) has 14$\%$ of strontium vacancies (marked V$_\mathrm{Sr}$), the TiO$_2$ termination (right) has 14$\%$ of Sr adatoms (Sr$_\mathrm{ad}$).   
		b,c) Changes in the work function measured by KPFM at $T=78$~K. On the SrO termination (b), the work function decreases and the change persists after switching off the illumination. The original work function is recovered after scanning the area by STM with a positive sample bias. For the TiO$_2$ termination (c), the UV light has no measurable effect on the work function. 
		d-g) Sequence of STM (d) and ncAFM images (e-g) demonstrating the spatial confinement of the work function modifications on the SrO termination. d) STM image of an area before the UV irradiation. e) AFM image of the same area after the UV illumination. Thereafter, the red square marked '1' was scanned in constant height using $V_\mathrm{S}=+1.7$~V, $I_\mathrm{T}\in [0;2.3]$~pA  and the square marked '2' was subsequently imaged by AFM, see panel (f). The area pre-scanned by STM appreas darker because of a different work function. The whole area was subsequently scanned in STM mode ($V_\mathrm{S}=+3.2$~V, $I_\mathrm{T}=15$~pA) and imaged by AFM, see panel (g). The whole image area was discharged, therefore the darker square in area '1' has vanished. 
	}\label{fig1}
\end{figure*}



SrTiO$_3$ is a prototype for ABO$_3$ cubic perovskite oxides. Each Sr atom is centred in a surrounding O cuboctahedron and Ti atoms are centred in O octahedrons \cite{lytle1964x}, resulting in a centrosymmetric cubic unit cell. The material became widely known for its electronic properties, such as the high dielectric permittivity  \cite{STOpermittivity1, STOpermittivity2} or two dimensional electron gases \cite{ohtomo2004high, santander-syro2011two, Wang2014} that form at its surfaces and interfaces. At the same time, SrTiO$_3$ is an inexpensive, stable, and non-toxic semiconductor \cite{phoon2019review} with an indirect band gap of 3.25~eV \cite{van2001bulk}, which makes it a candidate for charge separation in various photoelectronic and photocatalytic applications \cite{zhang2014new}. Doped SrTiO$_3$ achieves record efficiencies in photocatalytic water splitting \cite{STOphotocatalytic2023, domen1986mechanism,brookes1987srtio,liu2008synthesis, guan2014cocatalytic} and is frequently used for antibiotic photodegradation \cite{kumar2018high} or gas-phase NO removal \cite{zhang2016visible}. 

The behaviour of excess charges in complex ionic materials attracts attention from both the experimental and theoretical communities \cite{Sreekumar2025, Thornton2016, Wu2023, Fu2023, Redondo2024, Ellinger2023, PolaronReview2021, PolaronsReview2001, Michely2026}. Photoexcited charges can localize either at defects or self-localize in a perfect lattice, creating trapped charges or polarons, respectively. The lifetime of such quasiparticles may dramatically exceed the lifetime of delocalized carriers, resulting in effects that can be both advantagous or deleterious depending on each specific application. Here we demonstrate that the bulk-terminated SrTiO$_3$ surface prepared by cleaving can efficiently separate charge upon UV irradiation, creating photoexcited holes that become trapped at defects and show lifetimes exceeding a day at cryogenic temperatures. The single trapped holes are localized with atomic precision by a combined scanning tunneling microscopy/atomic force microscopy (STM/AFM) setup and the results are corroborated with density functional theory (DFT) calculations.   


All experiments were performed on a bulk-terminated SrTiO$_3$(001) surface prepared by strain-assisted \textit {in-situ} cleaving of bulk SrTiO$_3$ single crystals $n$-doped with 0.7 atomic percent (at.\%) Nb \cite{sokolovic2019incipient, SokolovicQuest2021, sokolovic2025cleave}. Experimental details are provided in electronic supplementary information (ESI) \cite{ESI}. Cleaving provides micrometer-size domains of SrO and TiO$_2$ terminations; atomic-scale details of these surfaces are shown in the constant-height AFM image in Fig.~\hyperref[fig1]{\ref{fig1}a}. The SrO termination (left) carries 14\% of defects, which were previously identified as strontium vacancies \cite{sokolovic2019incipient}, marked V$_\mathrm{Sr}$. The missing Sr atoms were left on the TiO$_2$ counterpart during the cleave, and they are present in the form of adatoms in the same concentration (Fig.~\hyperref[fig1]{\ref{fig1}a} right). The Sr vacancies and adatoms act as electron acceptors and donors, respectively, making the TiO$_2$ termination metallic \cite{sokolovic2019incipient,sokolovic2024duality}, while the SrO termination keeps the semiconducting character.   

Illuminating the entire cleaved surface by UV light (using an LED with $\lambda$=$365$\,nm) induces changes in the surface work function of the SrO termination, as shown by the Kelvin probe force measurements (KPFM) \cite{KPFM2012} in Fig.~\hyperref[fig1]{\ref{fig1}b}, while the metallic TiO$_2$ shows no changes (Fig.~\hyperref[fig1]{\ref{fig1}c}). During the UV irradiation, the local contact potential difference (LCPD) of the SrO termination changes from $-0.91$~V to $-1.81$~V. After turing the UV light off, the LCPD does not revert back, but stabilizes at $-1.72$~V, close to the value measured under the UV illumination. This value is stable over extended periods of time (days at $T$=5 or 78~K, see section SM7 \cite{ESI}), but the LCPD can be reverted back towards the original value by scanning the region in STM mode (here using sample bias, $V_\mathrm{S} = +3.2$V, tunneling current, $I_\mathrm{T} = 15$~pA, thus injecting electrons). In the text below, the LCPD values are converted into absolute surface work function $\phi$, using the fact that the tip was prepared and calibrated on a Cu(110) surface, here achieving $\phi_\mathrm{tip}$$\approx $4.6\,eV (see section SM3 \cite{ESI}). 

The entire SrO termination changes its work function after UV irradiation, but its discharging shows a strong spatial confinement, as shown in Fig.\,\hyperref[fig1]{\ref{fig1}d-g}. Panel d shows a region of the SrO termination imaged by STM. The surface was subsequently irradiated by UV light and the same region was imaged in the AFM mode; KPFM shows that the work function has decreased from 3.6~eV to 2.8~eV. A negative V$_\mathrm{S}$ of $-0.5$~V was used during the scan, which is necessary to avoid modification of the work function by injecting electrons into the surface. The region '1' marked by a red dashed square in Fig.\,\hyperref[fig1]{\ref{fig1}e} was subsequently imaged in STM mode (V$_\mathrm{S} = +3.2$~V, I$_\mathrm{T} = 15$~pA) and imaged by AFM, see Fig.\,\hyperref[fig1]{\ref{fig1}f}. The region pre-scanned by STM shows a distinctly different contrast in the AFM image, which is a fingerprint of a different work function in that region (decreased tip-sample attraction at V$_\mathrm{S} = 0$~V ; Kelvin parabolas measured inside and outside the square reveal $\phi_\mathrm{in} \approx 3.6$~eV and $\phi_\mathrm{out} \approx 2.8$~eV, respectively. Thereafter, the entire area marked as region `2' can be discharged by scanning in STM mode; an AFM image measured afterwards is shown in Fig.\,\hyperref[fig1]{\ref{fig1}g} where the darker square is gone due to a resulting uniform LCPD. 

These photoinduced work function changes therefore show a strong spatial confinement, persistence in time, and they are also reversible and can be repeated multiple times, see Fig.\,\hyperref[fig2]{\ref{fig2}a}. In principle, such effects could be either caused by formation of defects in the material, or by charge localization near the surface. The reversibility of the process clearly points at a model based on charge trapping and band bending, see Fig.\,\hyperref[fig2]{\ref{fig2}b}. The bulk material is an $n$-type semiconductor due to niobium doping (0.7~at.~$\%$). The TiO$_2$ termination contains Sr adatoms, which act as $n$-type dopants and induce downwards band bending, while the SrO termination contains Sr vacancies, which are $p$-type dopants and induce upwards band bending. The band-bending on the TiO$_2$ side is weak, since the Fermi level is already close to the conduction band minimum. The band bending at the SrO termination is substantially stronger and creates an internal electric field in the near-surface region. This efficiently separates photogenerated $e$-$h$ pairs when the sample is illuminated by UV light; photogenerated electrons are accelerated towards the bulk, while holes travel towards the surface. 

The holes accumulating at the SrO termination eventually drive the system to a flat-band condition; the work functions of the SrO and TiO$_2$ terminations become almost equal under UV irradiation (Fig.\,\hyperref[fig2]{\ref{fig2}a}). The initial upward band bending at the SrO termination is approximately 1 eV, and the UV reduces it approximately by 0.8 eV (Fig.\,\hyperref[fig2]{\ref{fig2}a}). The value of 0.8~eV was typically measured when the sample was kept at $T=78$~K, while measurements at $T=4.8$~K systematically provided a slightly lower value of 0.6~eV, see Section SM6 \cite{ESI}. These values provide a resonable match with electrostatic models. The depth of the depleted region $x_\mathrm{d}$ induced by the surface Sr vacancies is 

\begin{displaymath}
x_\mathrm{d}=\dfrac{c_\mathrm{VSr} \cdot q_\mathrm{VSr}}{c_\mathrm{Nb} \cdot q_\mathrm{Nb}},
\end{displaymath}

where the areal density of strontium vacancies $c_\mathrm{VSr}$ corresponds to 14$\%$ of surface unit cells, and we consider their charge state $q_\mathrm{VSr}$ to be $2e$. The bulk density of niobium dopants $c_\mathrm{Nb}$ corresponds to 0.7$\%$ of the unit cells, with a charge state of $q_\mathrm{Nb}= 1e$. This provides $x_\mathrm{d}$ of 15.6~nm (40 unit cells). The associated band bending is then

\begin{displaymath}
V_\mathrm{BB} = \int_{0}^{x_\mathrm{d}} E \,dx = \frac{1}{2}E_\mathrm{surf} \cdot x_\mathrm{d} = \frac{c_\mathrm{VSr} \cdot q_\mathrm{VSr}}{2\epsilon_0\epsilon_\mathrm{r}}x_\mathrm{d},
\end{displaymath}

where $E_\mathrm{surf}$ is the maximum electric field at the surface. The relative permittivity $\epsilon_\mathrm{r}$ of SrTiO$_3$ is 300 at room temperature and increases up to 10$^4$ at cryogenic temperatures \cite{STOpermittivity1}. However, it decreases substantially under high electric fields \cite{STOpermittivity2}. Taking $\epsilon_\mathrm{r}=300$ as a realistic effective value, we obtain $V_\mathrm{BB}=0.87$~eV, which is consistent with the experimental observations in Fig.\,\hyperref[fig2]{\ref{fig2}a}.

\begin{figure}[t]
		\includegraphics[width=1.0\columnwidth]{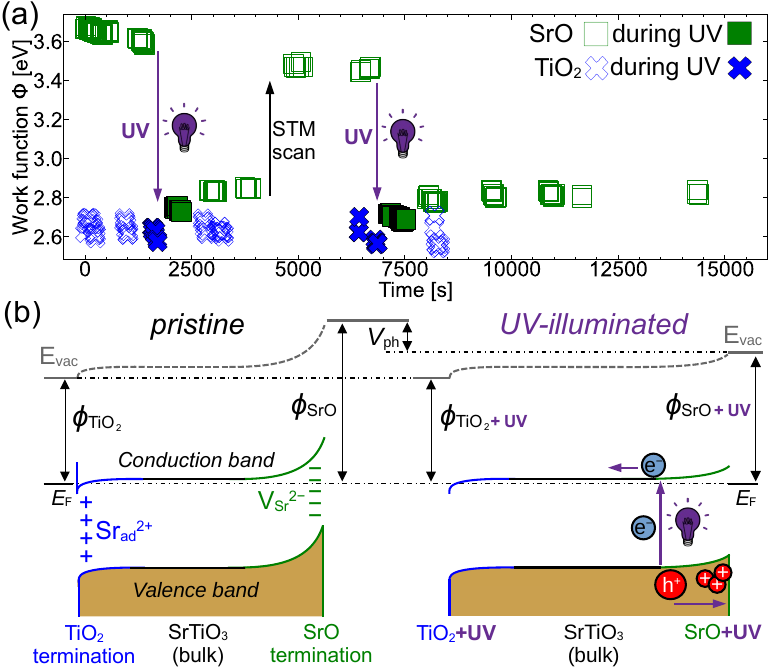}
	\caption{			
		\textbf{Model based on trapping of photoexcited holes.} 
		a) KPFM measurement of time-evolution of the work function $\Phi$ during several hours of experiment. The UV illumination changes the work function on the SrO termination (at $t\approx2000$~s), and the effect can be reverted by empty-state STM scan (at $t\approx4500$~s). The process is reversible (second UV illumination at $t\approx7000$~s). 
		b) Model of band structure for the pristine and the UV-irradiated surfaces.
	}\label{fig2}
\end{figure}

\begin{figure}
	\begin{center}
		\includegraphics[width=1.0\columnwidth,clip=true]{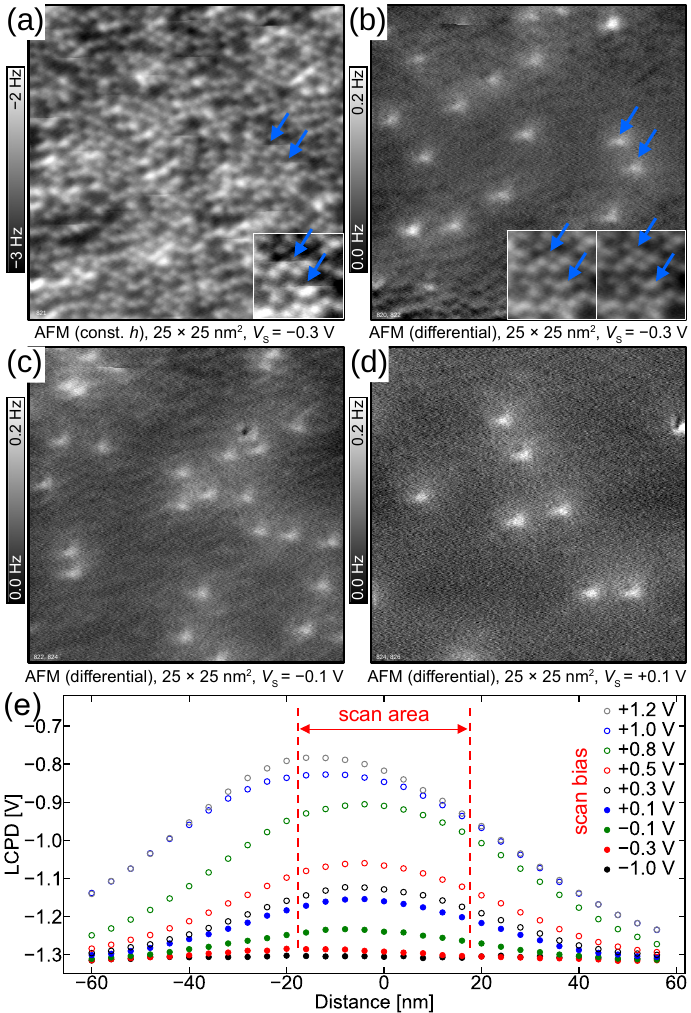}
	\end{center}
	\caption{			
		\textbf{Imaging single photoexcited holes.} a) Constant-height AFM image of a SrO-terminated surface illuminated by UV light. Horizontal streaks are attributed to elimination of photoexcited holes; two streaks are marked by arrows and shown in the inset with enhanced contrast. b) Difference of two AFM images measured after and before the intrusive scanning in panel a (see the text), showing the spatial distribution of electrostatic forces associated with the annihilated holes. The inset shows magnified AFM images near the two features marked by arrows, before (left) and after (right) imaging at V$_\mathrm{S}=-0.3$~V. c,d) Difference images similar to panel (b), showing the spatial localization of holes annihilated by subsequent scanning at -0.1 and +0.1~V, respectively. e) Evolution of the local contact potential when scanning the surface at gradually increasing bias. A central square region of 35$\times$35 nm$^2$ was scanned; the LCPD profiles were measured along a line 120 nm long. Panels a-d were measured at 4.5~K, panel e) at 78~K. Oscillation amplitude was 0.5 nm in all cases.
	}\label{fig3}
\end{figure}

Our experiments show that the surface keeps photoexcited holes over long time periods without recombination, while the local erasure of the charge in Figure~\hyperref[fig1]{\ref{fig1}f} also implies that the holes are immobile in the lateral direction. This is in principle possible for polarons or charges trapped at defects \cite{PolaronReview2021, PolaronsReview2001, cui2014surface}. Figure~\hyperref[fig3]{\ref{fig3}} shows how these localized charges are visualized in real space, based on the unique ability of nc-AFM to detect electrostatic forces with a sensitivity corresponding to single-electron charges \cite{Gross2009ScienceAu, Redondo2024, Fatayer2018, Berger2020}. Fig.~\hyperref[fig3]{\ref{fig3}a} shows a constant-height AFM image, where a SrO surface was illuminated by UV light and subsequently imaged at $V_\mathrm{S}=-0.3$~V. The tip-sample distance was approximately 1.2~nm, about 0.6 nm higher than typical conditions used for obtaining atomic resolution \cite{Ellner2016}; here the Sr vacancies dominate the image contrast. 
Figure~\hyperref[fig3]{\ref{fig3}a} contains horizontal streaks aligned with the scan direction, each with a length of $\approx$3 nm. A total of 22 such streaks are observed, two of which are marked by arrows.
We attribute these streaks to abrupt changes in the electrostatic force, caused by tip-induced elimination of the photoexcited holes.

The UV-irradiated surface could be imaged non-intrusively using a sample bias lower than $-0.5$~V. Raising the voltage above this threshold resulted in streaks such as in Fig.~\hyperref[fig3]{\ref{fig3}a}; further scans at the same voltage do not induce more changes. Additional details are provided in Section SM4 \cite{ESI}. Raising the bias towards more positive voltage induces new streaks. The following procedure was employed to visualize the holes in real space: The region was first imaged at $-0.5$~V, $i.e.$, non-intrusively, followed by scanning at a higher bias, and then again scanned at $-0.5$~V. The first and the last images were subtracted. The results are shown Fig. \hyperref[fig3]{\ref{fig3}b,c,d} for tip-induced changes caused by scanning at $V_\mathrm{S}=-0.3$, $-0.1$ and $+0.1$~V, respectively. Most of the features in Fig.~\hyperref[fig3]{\ref{fig3}b,c,d} have a similar shape and brightness. The features are predominantly centered at or near the Sr vacancies. The inset in Fig.~\hyperref[fig3]{\ref{fig3}b} shows AFM images of the two features marked by arrows, before and after the electron injections. The brightness of the two vacancies marked by arrows increases.

At first glance, it may seem surprising that the threshold bias for annihilating the holes is negative (V$_\mathrm{S}\approx-0.5$~V), while it requires electron injection from the tip. We link this to the fact that the UV-illuminated surface is in a non-equilibrium state. The photoinduced holes are 1 to 2~eV below the conduction band of the SrTiO$_3$ (Fig.~\hyperref[fig4]{\ref{fig4}}). Due to the downward band bending (Fig.~\hyperref[fig2]{\ref{fig2}b}), the holes can also be filled when the Fermi level of the tip is below the equilibrium Fermi level of the bulk SrTiO$_3$. For a complete picture, Fig.~\hyperref[fig3]{\ref{fig3}e} shows the evolution of the LCPD during the gradual elimination of the holes. Here, a larger area of 35$\times$35 nm$^2$ was scanned at gradually increasing voltages in the same way as in Fig.~\hyperref[fig3]{\ref{fig3}a-d} and the LCPD was measured across this reagion after each scan. Raw data for this experiment are provided in the ESI \cite{ESI}. 

It is noteworthy that the number of electron-injection events observed in Fig.~\hyperref[fig3]{\ref{fig3}} is significantly lower than expected from electrostatic considerations. While measuring the data in panels (b-d), we observed a total of 68 events and the LCPD has shifted by 0.2 eV, about a third of the total shift of $\approx$0.6~eV. The area contains approximately 500 Sr vacancies, therefore the expected number of holes is 500 to 1000, depending on the vacancy charge state.

\begin{figure}
	\begin{center}
		\includegraphics[width=1.0\columnwidth,clip=true]{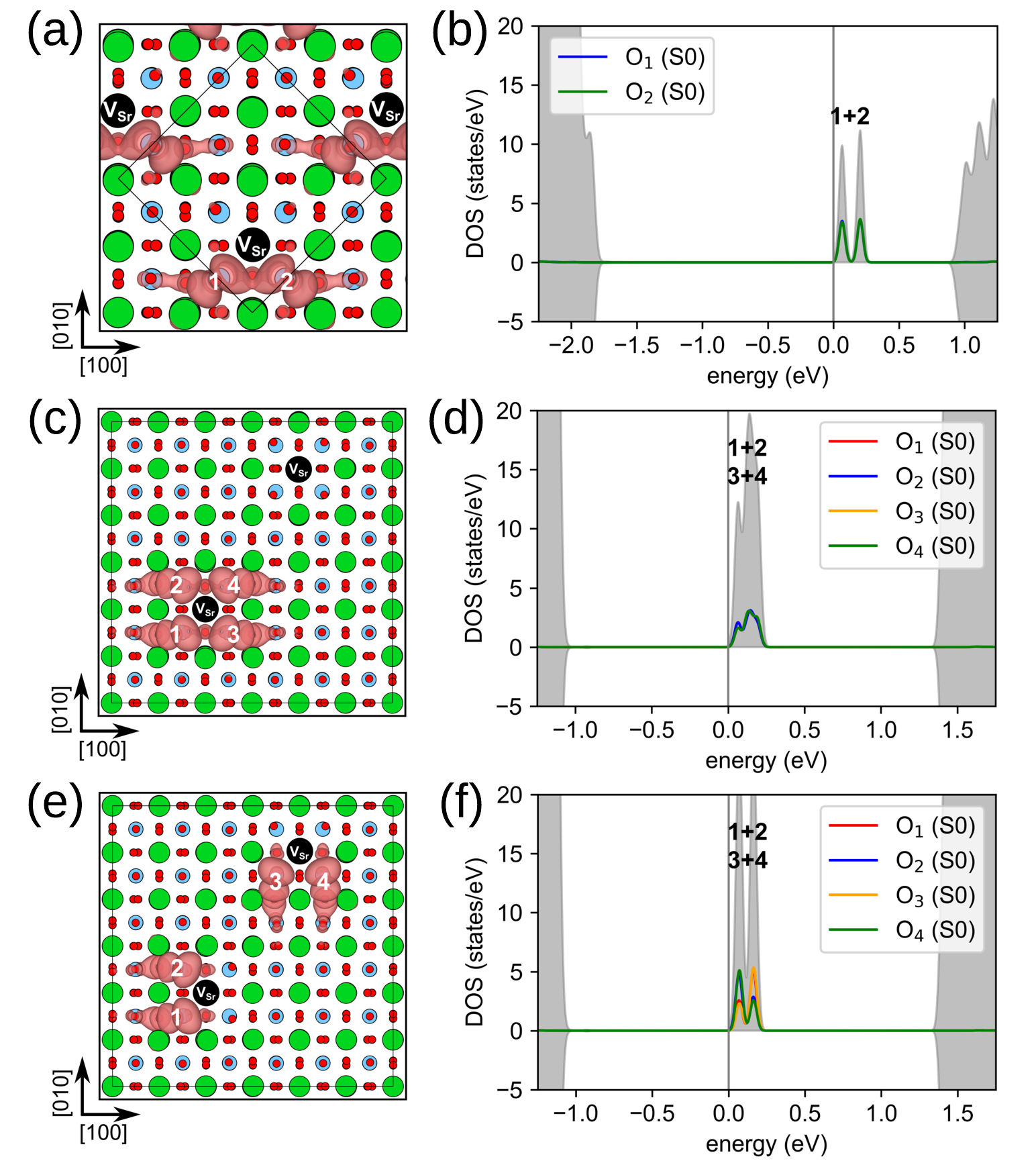}
	\end{center}
	\caption{			
		\textbf{Theoretical modelling of hole trapping} a) Isodensity surfaces of the gap states in $2\sqrt{2}\times2\sqrt{2}$ slab with one Sr vacancy. Two holes localize near the vacancy. b) The corresponding density of states showing the two-hole complex forming in the oxygen $p_\mathrm{O}$ surface states (S0). c-f) Analogous calculations performed on a larger $6\times6$ slab with two Sr vacancies. All four holes may localize at a single Sr vacancy (c,d) or acquire a slightly less favorable configuration where a pair is localized at each vacancy (e,f). 
	}\label{fig4}
\end{figure}

Theoretical modeling using slabs of different sizes was used to elucidate the mechanism of charge localization, see Fig. \hyperref[fig4]{\ref{fig4}}. In agreement with previous work \cite{Hao2015, Ellinger2023}, polaron formation was found neither on the surface, nor in the bulk region of a defect-free slab under equilibrium conditions.
However, adding positive charge by removing Sr species enables hole trapping in the planar $p_x$ and $p_y$ orbitals of surface oxygen atoms adjacent to the Sr vacancy. Figure~\hyperref[fig4]{\ref{fig4}a} contains one vacancy per a $2\sqrt{2}\times2\sqrt{2}$ unit cell, approximately corresponding to the experimentally measured vacancy concentration (12.5\% vs. 14\%). Hole localization next to V$_\mathrm{Sr}$ is favored by the electrostatic attraction between the hole and the negatively charged defect, as well as by the increased lattice flexibility that allows for local, polaron like, lattice distortion around the hole. The two-hole states exhibit a two-level peak structure in the density of states (DOS, Fig.~\hyperref[fig4]{\ref{fig4}b}), suggestive of a hybridized complex. Importantly, the trapped holes are highly stable, with a binding energy exceeding 200 meV, and localize exclusively in the surface layer.

Interaction between multiple defects was investigated using a larger $6\times6$ supercell with two Sr vacancies (providing four holes) see Fig. \hyperref[fig4]{\ref{fig4}c,e}. The clear tendency toward localization at the V$_\mathrm{Sr}$ sites persists, in agreement with experiment. Accumulation of all four holes adjacent to a single Sr vacancy is nearly isoenergetic (within 4 meV per hole) with configurations where the holes are distributed between two distant defect sites (Fig. \hyperref[fig4]{\ref{fig4}}c,e, respectively). The corresponding DOS shows well-defined in-gap states with increased energy spread (Fig. \hyperref[fig4]{\ref{fig4}}d,f). Even though the four-hole configuration in Fig. \hyperref[fig4]{\ref{fig4}c} nominally overcompensates the Sr vacancy, hybridization and clustering provide a stabilizing effect, favoring the formation of spatially localized multi-hole complexes.


The predicted absence of polaron formation in a defect-free lattice is in-line with the experimental observations, since all the tip-induced discharging events Fig.~\hyperref[fig3]{\ref{fig3}} are localized near the Sr vacancies. The calculations indicate that a single vacancy can hold one, two, or even more holes, and the states of adjacent holes hybridize. While it would be intuitive to assign the tip-induced discharging events in Fig.~\hyperref[fig2]{\ref{fig2}b-d} to single-hole events, the calculations indicate that each of them might be linked with a simultaneous annihilation of two, or even more holes: If there is a stable multi-hole complex and one hole is eliminated by tunneling, then the remaining holes may become destabilized and tunnel shortly after the initial event. This multi-hole scenario could explain the quantitative mismatch in the electrostatic considerations discussed above, where the number of experimentally detected discharging events is significantly lower than the number of charges expected at the surface. Another possible scenario could be the presence of subsurface trapping centres: Structural defects and impurities in the depletion layer could trap holes, where they cannot be easily visualized by AFM. Additional evidence for this picture is provided in Section SM5 \cite{ESI}, where a measurable tunneling current needs to be injected into the conduction band to recover the original surface work function.

The existence of long-lived trapped holes on the SrTiO$_3$(001) surface is surprising, considering the metallic character of the Nb-doped substrate. The lifetime of these charge carriers exceeds a day, which means that once the material is irradiated by UV light, electrons or X-rays, it can remain in a nonequilibrium state for any following experiments, such as photoemission \cite{sokolovic2024duality} or transport measurements. The longevity of these photoexcited states is possible due to the thickness of the depleted region ($\approx$15~nm), which excludes any tunneling-mediated recombination, combined with a low lateral mobility of the holes that could allow diffusion to defective surface regions (hot spots). Our work shows that the noncontact AFM holds a strong potential for probing and understanding excited electronic states and the methodology presented here is applicable to other systems as well.

This work was supported by the Czech Science Foundation, project GACR 20-21727X (M. Se., A. A., D. W. and L. A.) and the Ministry of Sports and Education (MSMT), project ERC CZ “PoTr” LL2324 (M.Se., A.A, L.A., and S. S.). The work was supported in part by the Austrian Science Fund (FWW), projects SuPer (P32148-N36) and SFB TACO (Grant DOI 10.55776/F8100). For open access purposes, the author has applied a CC BY public copyright license to any author-accepted manuscript version arising from this submission. The Joint Austrian (BMBWF CZ15/2021) and Czech (MSMT8J21AT004) project (M.R., M.Se., and F.E.) supported expenses for traveling. D. W. acknowledges the support from the Polish National Science Centre, project SONATA 2022/47/D/ST5/02439.
M. Se and Sr. S. acknowledge the support from the European Research Council (ERC) under the European Union’s Horizon Europe research and innovation programme (grant agreement No. 101169782, Consolidator Grant ‘SPOT’).

\bibliography{Bibliography5.bib}

@article{Hao2015,
  title = {Coexistence of trapped and free excess electrons in {SrTiO}$_3$},
  author = {Hao, Xianfeng and Wang, Zhiming and Schmid, Michael and Diebold, Ulrike and Franchini, Cesare},
  journal = {Phys. Rev. B},
  volume = {91},
  issue = {8},
  pages = {085204},
  numpages = {6},
  year = {2015},
  month = {Feb},
  publisher = {American Physical Society},
  doi = {10.1103/PhysRevB.91.085204},
  url = {https://link.aps.org/doi/10.1103/PhysRevB.91.085204}
}

@article{Ellinger2023,
  title = {Small polaron formation on the {N}b-doped {SrTiO}$_{3}$(001) surface},
  author = {Ellinger, Florian and Shafiq, Muhammad and Ahmad, Iftikhar and Reticcioli, Michele and Franchini, Cesare},
  journal = {Phys. Rev. Mater.},
  volume = {7},
  issue = {6},
  pages = {064602},
  numpages = {9},
  year = {2023},
  month = {Jun},
  publisher = {American Physical Society},
  doi = {10.1103/PhysRevMaterials.7.064602},
  url = {https://link.aps.org/doi/10.1103/PhysRevMaterials.7.064602}
}

@article{Wang2014,
author = {Zhiming Wang  and Zhicheng Zhong  and Xianfeng Hao  and Stefan Gerhold  and Bernhard Stöger  and Michael Schmid  and Jaime Sánchez-Barriga  and Andrei Varykhalov  and Cesare Franchini  and Karsten Held  and Ulrike Diebold },
title = {Anisotropic two-dimensional electron gas at {SrTiO}$_3$(110)},
journal = {Proc. Natl. Acad. Sci. USA},
volume = {111},
number = {11},
pages = {3933-3937},
year = {2014},
doi = {10.1073/pnas.1318304111}}

@article{VASP1,
    title = {{Efficient iterative schemes for ab initio total-energy calculations using a plane-wave basis set}},
    year = {1996},
    journal = {Phys. Rev. B},
    author = {Kresse, G. and Furthm{\"{u}}ller, J.},
    number = {16},
    month = {10},
    pages = {11169--11186},
    volume = {54},
    url = {https://link.aps.org/doi/10.1103/PhysRevB.54.11169},
    doi = {10.1103/PhysRevB.54.11169},
    issn = {0163-1829}
}

@article{VASP2,
    title = {{From ultrasoft pseudopotentials to the projector augmented-wave method}},
    year = {1999},
    journal = {Phys. Rev. B},
    author = {Kresse, G. and Joubert, D.},
    number = {3},
    month = {1},
    pages = {1758--1775},
    volume = {59},
    url = {https://link.aps.org/doi/10.1103/PhysRevB.59.1758},
    doi = {10.1103/PhysRevB.59.1758},
    issn = {0163-1829}
}

@article{Perdew1996a,
    title = {{Generalized gradient approximation made simple}},
    year = {1996},
    journal = {Phys. Rev. Lett.},
    author = {Perdew, John P. and Burke, Kieron and Ernzerhof, Matthias},
    number = {18},
    pages = {3865--3868},
    volume = {77},
    url = {https://link.aps.org/doi/10.1103/PhysRevLett.77.3865},
    isbn = {9780596529321},
    doi = {10.1103/PhysRevLett.77.3865},
    issn = {10797114},
    pmid = {10062328},
    arxivId = {10.1016/0927-0256(96)00008}
}

@article{Dudarev1998,
    title = {{Electron-energy-loss spectra and the structural stability of nickel oxide: An LSDA+U study}},
    year = {1998},
    journal = {Phys. Rev. B},
    author = {Dudarev, S. and Botton, G.},
    number = {3},
    month = {1},
    pages = {1505--1509},
    volume = {57},
    url = {https://link.aps.org/doi/10.1103/PhysRevB.57.1505},
    isbn = {0163-1829},
    doi = {10.1103/PhysRevB.57.1505},
    issn = {1550235X}
}

@article{Allen2014,
    title = {{Occupation matrix control of d- and f-electron localisations using DFT + U}},
    year = {2014},
    journal = {Phys. Chem. Chem. Phys.},
    author = {Allen, Jeremy P. and Watson, Graeme W.},
    number = {39},
    pages = {21016--21031},
    volume = {16},
    publisher = {Royal Society of Chemistry},
    url = {http://xlink.rsc.org/?DOI=C4CP01083C},
    doi = {10.1039/c4cp01083c},
    issn = {14639076},
    pmid = {24832683}
}

@article{ldipolvasp,
  title = {Adsorbate-substrate and adsorbate-adsorbate interactions of {N}a and {K} adlayers on {A}l(111)},
  author = {Neugebauer, J\"org and Scheffler, Matthias},
  journal = {Phys. Rev. B},
  volume = {46},
  issue = {24},
  pages = {16067--16080},
  numpages = {0},
  year = {1992},
  month = {Dec},
  publisher = {American Physical Society},
  doi = {10.1103/PhysRevB.46.16067},
  url = {https://link.aps.org/doi/10.1103/PhysRevB.46.16067}
}

@misc{Yalcin2025arxiv,
   author = {Firat Yalcin and Carla Verdi and Viktor C. Birschitzky and Matthias Meier and Michael Wolloch and Michele Reticcioli},
   title = {Automated Modeling of Polarons: Defects and Reactivity on {TiO}$_2$(110) Surfaces},
   journal = {Npj. Comput. Mater.},
   volume = {10},   
   pages = {89},
   year = {2024},
}

@article{Birschitzky2024,
author = {Birschitzky, Viktor C. and Sokolovi{\'{c}}, Igor and Prezzi, Michael and Palot{\'{a}}s, Kriszti{\'{a}}n and Setv{\'{i}}n, Martin and Diebold, Ulrike and Reticcioli, Michele and Franchini, Cesare},
doi = {10.1038/s41524-024-01289-4},
issn = {2057-3960},
journal = {npj Computational Materials},
month = {may},
number = {1},
pages = {89},
title = {{Machine learning-based prediction of polaron-vacancy patterns on the TiO2(110) surface}},
url = {https://www.nature.com/articles/s41524-024-01289-4},
volume = {10},
year = {2024},
archivePrefix = {arXiv},
arxivId = {2401.12042},
eprint = {2401.12042}
}

@article{domen1986mechanism,
  title={Mechanism of photocatalytic decomposition of water into {H}$_2$ and {O}$_2$ over {NiO}--{SrTiO}$_3$},
  author={Domen, Kazunari and Kudo, Akihiko and Onishi, Takaharu},
  journal={J. Catal.},
  volume={102},
  number={1},
  pages={92--98},
  year={1986},
  publisher={Elsevier}
}

@article{ohtomo2004high,
  title={A high-mobility electron gas at the {LaAlO}$_3$/{SrTiO}$_3$ heterointerface},
  author={Ohtomo, A and Hwang, H Y},
  journal={Nature},
  volume={427},
  number={6973},
  pages={423},
  year={2004},
  publisher={Nature Publishing Group}
}

@article{santander-syro2011two,
  title={Two-dimensional electron gas with universal subbands at the surface of {SrTiO}$_3$},
  author={Santander-Syro, A F and Copie, O and Kondo, T and Fortuna, F and Pailhes, S and Weht, R and Qiu, X G and Bertran, F and Nicolaou, A and Taleb-Ibrahimi, A and Le Fèvre, P and Herranz, G and Bibes, M and Reyren, N and Apertet, Y and Lecoeur, P and  Barthélémy, A and Rozenberg, M J},
  journal={Nature},
  volume={469},
  number={7329},
  pages={189},
  year={2011},
  publisher={Nature Publishing Group}
}

@article{sokolovic2019incipient,
  title={Incipient ferroelectricity: A route towards bulk-terminated {SrTiO}$_3$},
  author={Sokolovi{\'c}, Igor and Schmid, Michael and Diebold, Ulrike and Setv{\'\i}n, Martin},
  journal={Phys. Rev. Mater.},
  volume={3},
  number={3},
  pages={034407},
  year={2019},
  publisher={APS}
}

@misc{giessibl2013sensor,
  title={Sensor for noncontact profiling of a surface},
  author={Giessibl, Franz Josef},
  year={2013},
  month=mar # "~5",
  publisher={Google Patents},
  note={ {U}S Patent 8,393,009}
}

@article{setvin2012ultrasharp,
  title={Ultrasharp tungsten tips—characterization and nondestructive cleaning},
  author={Setv{\'\i}n, M and Javorsk{\`y}, J and Tur{\v{c}}inkov{\'a}, D and Matol{\'\i}nov{\'a}, I and Sobot{\'\i}k, P and Koc{\'a}n, P and O{\v{s}}t’{\'a}dal, I},
  journal={Ultramicroscopy},
  volume={113},
  pages={152--157},
  year={2012},
  publisher={Elsevier}
}

@article{huber2017low,
  title={Low noise current preamplifier for q{P}lus sensor deflection signal detection in atomic force microscopy at room and low temperatures},
  author={Huber, Ferdinand and Giessibl, Franz J},
  journal={Rev. Sci. Instr.},
  volume={88},
  number={7},
  pages={073702},
  year={2017},
  publisher={AIP Publishing LLC}
}

@article{gartland1972photoelectric,
  title={Photoelectric work function of a copper single crystal for the (100), (110), (111), and (112) faces},
  author={Gartland, P.O. and Berge, S. and Slagsvold, B.J.},
  journal={Phys. Rev. Lett.},
  volume={28},
  pages={738},
  year={1972},
  publisher={APS}
}

@article{sokolovic2025cleave,
  title={How to cleave cubic perovskite oxides},
  author={Sokolovi{\'c}, I. and Schmid, M. and Diebold, U. and Setv{\'\i}n, M.},
  journal={Rev. Sci. Instr.},
  volume={96},
  number={3},
  year={2025},
  publisher={AIP Publishing}
}

@article{sokolovic2024duality,
  title={Duality and degeneracy lifting in two-dimensional electron liquids on {SrTiO}$_3$(001)},
  author={Sokolovi{\'c}, I. and Guedes, E. B. and van Waas, T. P. and Polley, C. and Schmid, M. and Diebold, U. and Radovi{\'c}, M. and Setv{\'\i}n, M. and Dil, J.H.},
  journal={Nat. Commun.},
  volume={16},
  pages={4594},
  year={2025}
}

@article{phoon2019review,
  title={A review of synthesis and morphology of {S}r{T}i{O}$_3$  for energy and other applications},
  author={Phoon, B. L. and Lai, C. W. and Juan, J. C. and Show, P.-L. and Chen, W.-H.},
  journal={Int. J. Energy Res.},
  volume={43},
  pages={5151},
  year={2019},
  publisher={Wiley Online Library}
}

@article{zhang2014new,
  title={New understanding of the difference of photocatalytic activity among anatase, rutile and brookite {T}i{O}$_2$},
  author={Zhang, J. and Zhou, P. and Liu, J. and Yu, J.},
  journal={Phys. Chem. Chem. Phys.},
  volume={16},
  pages={20382},
  year={2014},
  publisher={The Royal Society of Chemistry}
}

@article{liu2008synthesis,
  title={Synthesis and high photocatalytic hydrogen production of {S}r{T}i{O}$_3$ nanoparticles from water splitting under {UV} irradiation},
  author={Liu, Y. and Xie, L. and Li, Y. and Yang, R. and Qu, J. and Li, Y. and Li, X.},
  journal={J. Power Sources},
  volume={183},
  pages={701},
  year={2008},
  publisher={Elsevier}
}

@article{guan2014cocatalytic,
  title={Cocatalytic effect of {S}r{T}i{O}$_3$ on {A}g$_3${PO}$_4$ toward enhanced photocatalytic water oxidation},
  author={Guan, X. and Guo, L.},
  journal={ACS Catal.},
  volume={4},
  pages={3020},
  year={2014},
  publisher={ACS Publications}
}

@article{kumar2018high,
  title={High-performance photocatalytic hydrogen production and degradation of levofloxacin by wide spectrum-responsive {A}g/{F}e$_3${O}$_4$ bridged {S}r{T}i{O}$_3$/g-{C}$_3${N}$_4$ plasmonic nanojunctions: joint effect of {A}g and {F}e$_3${O}$_4$ },
  author={Kumar, A. and Rana, A. and Sharma, G. and Naushad, M. and Al-Muhtaseb, A. and Guo, C. and Iglesias-Juez, A. and Stadler, F. J.},
  journal={ACS Appl. Mater. Interfaces},
  volume={10},
  pages={40474},
  year={2018},
  publisher={ACS Publications}
}

@article{zhang2016visible,
  title={Visible-light-active plasmonic {A}g--{S}r{T}i{O}$_3$ nanocomposites for the degradation of {NO} in air with high selectivity},
  author={Zhang, Q. and Huang, Y. and Xu, L. and Cao, J. and Ho, W. and Lee, S. C.},
  journal={ACS Appl. Mater. Interfaces.},
  volume={8},
  pages={4165},
  year={2016},
  publisher={ACS Publications}
}

@article{brookes1987srtio,
  title={{SrTiO}$_3$(100) step sites as catalytic centers for {H}$_2${O} dissociation},
  author={Brookes, N. B. and Thornton, G. and Quinn, F. M.},
  journal={Solid State Commun.},
  volume={64},
  pages={383},
  year={1987},
  publisher={Elsevier}
}

@article{lytle1964x,
  title={X-ray diffractometry of low-temperature phase transformations in strontium titanate},
  author={Lytle, F. W. },
  journal={J. Appl. Phys.},
  volume={35},
  pages={2212},
  year={1964},
  publisher={American Institute of Physics}
}

@article{van2001bulk,
  title={Bulk electronic structure of {S}r{T}i{O}$_3$: {E}xperiment and theory},
  author={van Benthem, K. and Els{\"a}sser, C. and French, R. H.},
  journal={J. App. Phys.},
  volume={90},
  pages={6156},
  year={2001},
  publisher={American Institute of Physics}
}

@article{STOphotocatalytic2023,
  title={Photocatalytic Overall Water Splitting by SrTiO$_3$: Progress Report and Design Strategies},
  author={Avcioglu, C. and Avcioglu, S. and Bekheet, M. F. and Gurlo, A.},
  journal={ACS Appl. Energy Mat.},
  volume={6},
  pages={1134-1154},
  year={2023}
}

@article{PolaronReview2021,
  title={Polarons in Materials},
  author={Franchini, C. and Reticcioli M. and Setvin, M. and Diebold, U.},
  journal={Nat. Rev. Mater.},
  volume={6},
  pages={560-586},
  year={2021}
}

@article{SokolovicQuest2021,
  title={Quest for a pristine unreconstructed {S}r{T}i{O}$_3$ surface: An atomically resolved study via noncontact atomic force microscopy},
  author={Sokolovi{\'c}, I. and Franchesci, G. and Wang, Z. and Xu, J. and Pavelec, J. and Riva, M. and Schmid, M. and Diebold, U. and Setvin, M.},
  journal={Phys. Rev. B},
  volume={103},
  pages={L241406},
  year={2021}
}

@article{STOpermittivity1,
  title={Permittivity of Strontium Titanate},
  author={Neville, R. C. and Hoeneisen, B. and Mead, C. A.},
  journal={J. Appl. Phys.},
  volume={43},
  pages={2124-2131},
  year={1972}
}

@article{STOpermittivity2,
  title={Field Dependent Permittivity in Metal-Semiconducting {SrTiO}$_3$ {S}chottky Diodes},
  author={van der Berg, R. A. and Blom, P. W. M. and Cillessen, J. F. M. and Wolf, R. M.},
  journal={Appl. Phys. Lett.},
  volume={66},
  pages={697-699},
  year={1995}
}

@article{PolaronsReview2001,
  title={Polarons in Crystalline and Non-crystalline Materials},
  author={Austain, I. G. and Mott, N. F.},
  journal={Advances in Physics},
  volume={50},
  pages={757-812},
  year={2001}
}

@BOOK{KPFM2012,
   author = "S. Sadewasser and T. Glatzel",
   title = "Kelvin Probe Force Microscopy",
   publisher = "Springer-Verlag Berlin Heidelberg",
   year = "2012"
}

@ARTICLE{Gross2009ScienceAu,
  author = "L. Gross and F. Mohn and P. Liljeroth and F. J. Giessibl and G. Meyer",
   title = "Measuring the Charge State of an Adatom with Noncontact Atomic Force Microscopy",
   journal = "Science",
   volume = "324",
   pages = "1428-1431",
   year = "2009"
}

@ARTICLE{Fatayer2018,
  author = "S. Fatayer and B. Schuler and W. Steurer and I. Scivetti and J. Repp and L. Gross and M. Persson and G. Meyer",
   title = "Reorganization energy upon charging a single molecule on an insulator measured by atomic force microscopy",
   journal = "Nat. Nanotechnol.",
   volume = "13",
   pages = "376-380",
   year = "2018"
}

@ARTICLE{Berger2020,
  author = "J. Berger and M. Ondracek and O. Stetsovych and P. Malý and P. Holy and J. Rybacek and M. Svec and I. Stara and T. Mancal and I. Stary and P. Jelinek",
   title = "Quantum dissipation driven by electron transfer within a single molecule investigated with atomic force microscopy",
   journal = "Nat. Commun.",
   volume = "11",
   pages = "1337",
   year = "2020"
}

@ARTICLE{Redondo2024,
  author = "J. Redondo and M. Reticcioli and V. Gabriel and D. Wrana and F. Ellinger and M. Riva and G. Franceschi and E. Rheinfrank and I. Sokolovic and Z. Jakub and F. Kraushofer and A. Alexander and E. Belas and L. Patera and J. Repp and M. Schmid and U. Diebold and G. Parkinson and C. Franchini and P. Kocan and M. Setvin",
   title = "Real-space investigation of polarons in hematite {F}e$_2${O}$_3$",
   journal = "Sci. Adv.",
   volume = "10",
   pages = "eadp7833",
   year = "2024"
}

@ARTICLE{Sreekumar2025,
  author = "S. Sreekumar and P. Kocan and M. Setvin",
   title = "Tracking polarons in real space by {STM/AFM}",
   journal = "Appl. Phys. Lett.",
   volume = "10",
   pages = "140502",
   year = "2025"
}

@ARTICLE{Wu2023,
  author = "H. Liu and A. Wang and P. Zhang and Ch. Ma and C. Chen and Z. Liu and Y.-Q. Zhang and B. Feng and P. Cheng and J. Zhao and L. Chen and K. Wu",
   title = "Atomic-scale manipulation of single-polaron in a two-dimensional semiconductor",
   journal = "Nat. Commun.",
   volume = "14",
   pages = "3690",
   year = "2023"
}

@ARTICLE{Fu2023,
  author = "M. Cai and M.-P. Miao and Y. Liang and Z. Jiang and Z.-Y. Liu and W.-H. Zhang and X. Liao and L.-F. Zhu and Da. West and S. Zhang and Y.-S. Fu",
   title = "Manipulating single excess electrons in monolayer transition metal dihalide",
   journal = "Nat. Commun.",
   volume = "14",
   pages = "3691",
   year = "2023"
}

@ARTICLE{Thornton2016,
  author = "C.M. Yim and M.B. Watkins and M.J. Wolf and C.L. Pang and K. Hermansson and G. Thornton",
   title = "Engineering Polarons at a Metal Oxide Surface",
   journal = "Phys. Rev. Lett.",
   volume = "117",
   pages = "116402",
   year = "2016"
}

@ARTICLE{Michely2026,
  author = "A. Safeer and O. Guleryuz and G. Miao and W. Jolie and T. Michely and J. Fischer",
   title = "Substrate Role in Polaron Formation on Single-layer Transition Metal Dihalides",
   journal = "Arxiv preprint",
   pages = "https://arxiv.org/abs/2512.21163",
   year = "2025"
}

@misc{ESI,
   title = "Supplementary information available online",
}

@article{Ellner2016,
author = {Ellner, Michael and Pavliček, Niko and Pou, Pablo and Schuler, Bruno and Moll, Nikolaj and Meyer, Gerhard and Gross, Leo and Per{\'e}z, Rub{\'e}n},
title = {The Electric Field of CO Tips and Its Relevance for Atomic Force Microscopy},
journal = {Nano Letters},
volume = {16},
number = {3},
pages = {1974-1980},
year = {2016},
doi = {10.1021/acs.nanolett.5b05251},
    note ={PMID: 26840626},

URL = {https://doi.org/10.1021/acs.nanolett.5b05251},
eprint = {https://doi.org/10.1021/acs.nanolett.5b05251}
}

@article{cui2014surface,
  title={Surface defects and their impact on the electronic structure of Mo-doped CaO films: an STM and DFT study},
  author={Cui, Yi and Shao, Xiang and Prada, Stefano and Giordano, Livia and Pacchioni, Gianfranco and Freund, Hans-Joachim and Nilius, Niklas},
  journal={Physical Chemistry Chemical Physics},
  volume={16},
  number={25},
  pages={12764--12772},
  year={2014},
  publisher={Royal Society of Chemistry}
}

\pagebreak

\title{Supplementary Information: Photoexcited Hole States at the SrTiO$_3$(001) Surface Imaged with Noncontact AFM}

\author{Igor Sokolović}
\affiliation{Institute of Applied Physics, TU Wien, Wiedner Hauptstrasse 8-10/134, 1040 Vienna, Austria}

\author{Florian Ellinger}
\affiliation{University of Vienna, Faculty of Physics and Center for Computational Materials Science, Vienna, Austria}

\author{Aji Alexander}
\affiliation{Department of Surface and Plasma Science, Faculty of Mathematics and Physics, Charles University, 180 00 Prague 8, Czech Republic}

\author{Dominik Wrana}
\affiliation{Department of Surface and Plasma Science, Faculty of Mathematics and Physics, Charles University, 180 00 Prague 8, Czech Republic}
\affiliation{Marian Smoluchowski Institute of Physics, Jagiellonian University, Krakow 30-348, Poland}

\author{Llorenc Albons}
\affiliation{Department of Surface and Plasma Science, Faculty of Mathematics and Physics, Charles University, 180 00 Prague 8, Czech Republic}

\author{Sreehari Sreekumar}
\affiliation{Department of Surface and Plasma Science, Faculty of Mathematics and Physics, Charles University, 180 00 Prague 8, Czech Republic}

\author{Michael Schmid}
\affiliation{Institute of Applied Physics, TU Wien, Wiedner Hauptstrasse 8-10/134, 1040 Vienna, Austria}

\author{Ulrike Diebold}
\affiliation{Institute of Applied Physics, TU Wien, Wiedner Hauptstrasse 8-10/134, 1040 Vienna, Austria}

\author{Michele Reticcioli}
\affiliation{Institute of Applied Physics, TU Wien, Wiedner Hauptstrasse 8-10/134, 1040 Vienna, Austria}
\affiliation{University of Vienna, Faculty of Physics and Center for Computational Materials Science, Vienna, Austria}
\affiliation{CNR-SPIN, c/o Department of Physical and Chemical Sciences, University of L’Aquila, Via Vetoio, L’Aquila I-67100, Italy}

\author{Cesare Franchini}
\email[Corresponding author: ]{cesare.franchini@univie.ac.at}
\affiliation{University of Vienna, Faculty of Physics and Center for Computational Materials Science, Vienna, Austria}
\affiliation{Dipartimento di Fisica e Astronomia, Universit\`a di Bologna, 40127 Bologna, Italy}

\author{Martin Setvin}
\email[Corresponding author: ]{martin.setvin@matfyz.cuni.cz}
\affiliation{Department of Surface and Plasma Science, Faculty of Mathematics and Physics, Charles University, 180 00 Prague 8, Czech Republic}
\affiliation{Institute of Applied Physics, TU Wien, Wiedner Hauptstrasse 8-10/134, 1040 Vienna, Austria}


\maketitle

\tableofcontents

\section{Experimental Details}

\textbf{SrTiO$_3$ single crystals} were purchased from MaTeck GmbH in a custom shape suitable for cleaving. Samples were doped with 0.7~at\% of Nb and cut to slabs of approximately 2$\times$3$\times$8~mm$^3$, with the crystal orientation of 2$\times$3~mm$^2$ for the (001) face, and 3$\times$8~mm$^2$ for the (010) face. SrTiO$_3$(001) bulk-truncated surfaces were achieved by cleaving single crystals at room temperature in UHV chamber with base pressure lower than 1$\times$10$^{-10}$~mbar, using a stainless steel cleaving device \cite{sokolovic2019incipient,sokolovic2025cleave}, and $in-situ$ transferred into the cryogenic STM head within less than 4~min after the cleaving.

\textbf{ncAFM/STM measurements} were conducted in a vacuum chamber with a base pressure below 10$^{-10}$~mbar, using a ScientaOmicron qPlus STM/AFM head (LTSTM and POLAR systems with a similar configuration and technical parameters were used). ncAFM/STM was performed at cryogenic temperatures (LN$_2$ or LHe), using qPlus sensors with a separate wire for the tunneling current~\cite{giessibl2013sensor} and electrochemically etched W tips. The tips were self-sputtered in situ with Ar$^+$ ions~\cite{setvin2012ultrasharp}, and the apex was additionally shaped on Cu(110) to ensure its metallic character and well-defined work function. A cryogenic differential preamplifier~\cite{huber2017low} was used for deflection detection.

\textbf{UV irradiation} was generated with a 1~W UV LED with $\lambda=365$~nm, and focused onto a sample surface via an optical condenser from outside the vacuum chamber through a quartz window and openings in the cryostat shields. The spot size was circular with a diameter of $\approx$3~mm; the power at the surface is estimated as $\approx$0.1\,W, corresponding to a photon flux in order of 10$^{16}$~photons$\cdot$s$^{-1}$mm$^{-2}$. The photovoltage reached saturation after a few minutes of irradiaion; we typically applied the UV light for 10 to 30 minutes before performing the AFM experiments. Some experiments were performed with a Hg discharge UV lamp ($\lambda$=254\,nm), obtaining the same results. Prior to UV experiments, the whole measurement head was thoroughly irradiated with UV to induce photodesorption of molecules adsorbed on the cryostat walls and prevent contamination of the sample. Irradiation with the UV diode induced a slight temperature increase of less than 1\,K on the sample. During the UV irradiation, ncAFM/STM tip was kept in relative proximity to the surface ($\approx 100$\,nm away), by employing a constant-frequency-shift feedback loop.

\textbf{Kelvin parabolas} were performed by sweeping the sample bias while keeping the tip-sample separation constant, at sufficiently large distances ($\approx$5~nm away from the surface) to avoid electron tunneling. Efforts were performed to keep this tip-sample separation consistent as the Kelvin parabolas exhibit distance-dependent shifts in the order of $\approx$0.2~V. Conversion of CPD measurements to work function values were performed by calibrating the tip work function on a Cu(110) surface\cite{gartland1972photoelectric} before and after measuring on SrTiO$_3$(001), according to  $\varPhi[\mathrm{SrTiO_3(001)}]=\varPhi[\mathrm{Cu(110)}]-\mathrm{CPD}[\mathrm{Cu(110)]}-\mathrm{CPD}[\mathrm{SrTiO_3(001)}]$; we  overestimate  the $\varPhi[\mathrm{SrTiO_3(001)}]$ measurement error to $\pm0.2$~eV. More details on the work function evaluation procedure are thoroughly laid out in Sec.\,SM1

\raggedbottom
\pagebreak

\section{Computational Details}

\textbf{Density-Functional Theory} (DFT) calculations were performed using the Vienna ab-initio simulation package (VASP)~\cite{VASP1,VASP2}.
We adopted the Perdew, Burke, and Ernzerhof (PBE) parametrization~\cite{Perdew1996a}, with the inclusion of an on-site effective $U$~\cite{Dudarev1998} of $4.6$~eV and $5.25$~eV on the $d$ orbitals of Ti atoms and $p$ orbitals of O atoms, respectively.
We used an energy cutoff of $400$~eV and $\Gamma$-only sampling of the reciprocal space.
We constructed several SrTiO$_3$(001) slabs starting from the tetragonal bulk unit cell with lattice constants relaxed to minimize forces at PBE level ($a=3.958$~\AA, $c=3.986$~\AA):
$2\sqrt{2}\times2\sqrt{2}$ stoichiometric (asymmetric) slabs with 6 layers (2 layers fixed to bulk positions) were used to model a surface Sr vacancy coverage of 12.5\% by removing one surface Sr atom.
Larger $6\times6$ slabs were used to model a defect coverage of 5.6\%, by removing two Sr atoms from the surface layer.
Dipole corrections were applied to cancel residual electric fields in the vacuum~\cite{ldipolvasp}.
To prevent random electronic configurations~\cite{Birschitzky2024,  Yalcin2025arxiv}, hole localization at target sites was achieved using the occupation matrix control tool in the intermediate steps of the calculations~\cite{Allen2014}.
The spatial distribution of the charge densities was visualized by showing the in-gap hole states.

\raggedbottom

\section{Work functions measurements of the S{r}O and T{i}O$_2$ terminations} \label{sec:WF}

\begin{figure}[h!]
	\begin{center}
		\includegraphics[width=1\columnwidth,clip=true]{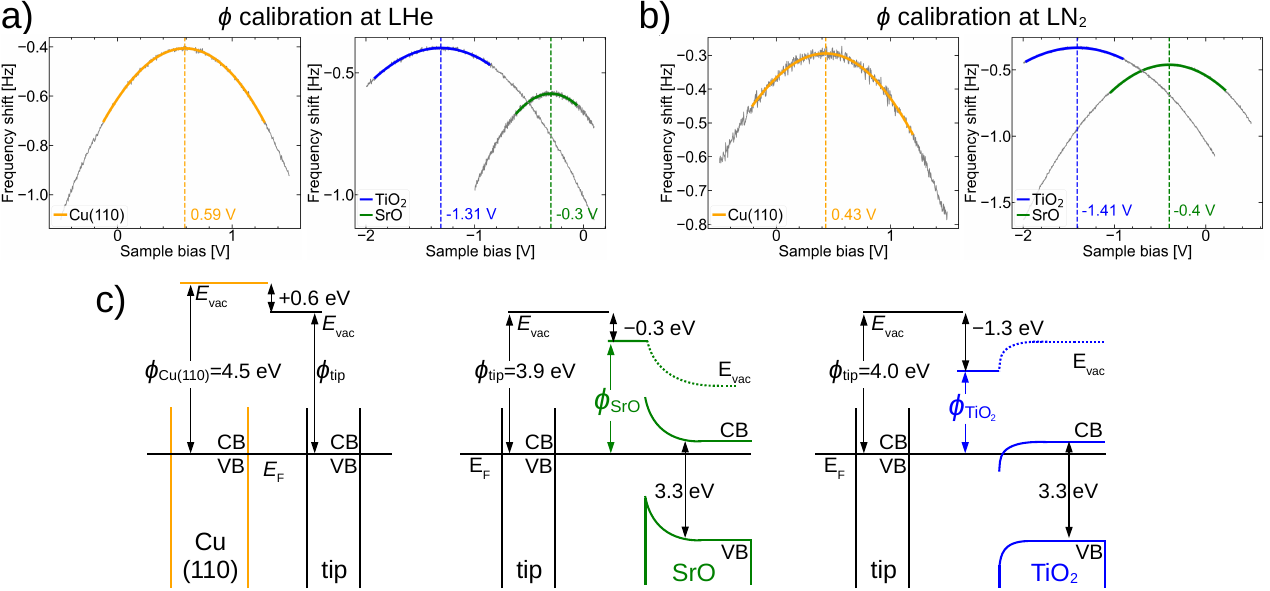}
	\end{center}
	\caption{ \textbf{Quantification of the work function of the SrO- and TiO$_2$-terminated SrTiO$_3$ by means of Kelvin probe force spectroscopy calibrated on a Cu(110) surface.} 
		(a) The maximum of the Kelvin parabola measured above the Cu(110) surface indicates a mismatch of vacuum levels between the atomically sharp ncAFM tip and the Cu(110) surface. The same tip was used to perform Kelvin parabolas on different surface terminations of freshly cleaved SrTiO$_3$(001), and determine their respective CPDs. Measurements above both Cu(110) and cleaved SrTiO$_3$ were performed at LHe (panel a) and LN$_2$ (panel b) temperatures. The work function difference between the two terminations of cleaved STO is $\approx$1\,eV in both cases.	
		c) Sketch (not to scale) illustrating the conversion of measured CPDs to work functions of (left-to-right): ncAFM tip based on a known Cu(110) work function, allows evaluating the work functions of the SrO and TiO$_2$ terminations.
	}\label{fig:WF}
\end{figure}

Kelvin probe force microscopy (KPFM) \cite{KPFM2012} allows to measure a relative difference of the local work function of the sample with respect to the tip apex. In order to obtain the absolute value of the surface work function, the tip work function was first callibrated on the Cu(110) surface and the same tip was used to characterize the SrTiO$_3$, with a special care to avoid any tip-sample interactions that could change the tip work function. The quantitative results are shown in Fig.~\ref{fig:WF}. The two experiments in Figs.\hyperref[fig:WF]{\ref{fig:WF}a}\,and\,\hyperref[fig:WF]{\ref{fig:WF}b} were identical in all aspects, but performed at two different temperatures. 

In both experiments in Figs.\hyperref[fig:WF]{\ref{fig:WF}a}\,and\,\hyperref[fig:WF]{\ref{fig:WF}b}
the CPD over a Cu(110) surface (CPD[tip:Cu(110)]) falls with a positive sample bias, indicating that the work function of the ncAFM tip $\phi$[tip] was lower than the work function of the Cu(110) surface $\phi$[Cu(110)], as illustrated in the sketch in Fig.\hyperref[fig:WF]{\ref{fig:WF}c}. In such case, a known reference value for $\phi$[Cu(110)] of 4.5\,eV \cite{gartland1972photoelectric} (rounded to the first decimal) can be used to determine $\phi$[tip] as:
\begin{equation*}
	\phi[\mathrm{tip}]=\phi[\mathrm{Cu(110)}]-\mathrm{CPD}[\mathrm{tip:Cu(110)}]
\end{equation*} 
For the experiment at LN$_2$ temperature [Fig.\hyperref[fig:WF]{\ref{fig:WF}a}] the tip work function measures $\phi$[tip]=4.5\,eV--0.6\,eV=3.9\,eV, while for the LHe experiment [Fig.\hyperref[fig:WF]{\ref{fig:WF}b}] it measures $\phi$[tip]=4.5\,eV--0.4\,eV=4.1\,eV.

On both surface terminations of cleaved SrTiO$_3$ such ncAFM tips measured CPD in the negative sample bias range, indicating that their work functions $\phi$[SrO] and $\phi$[TiO$_2$] are smaller than the work function of the tip. Their respective work functions can be determined based on the calibrated $\phi$[tip] as:
\begin{equation*}
	\phi[\mathrm{SrO}]=\phi[\mathrm{Cu(110)}]-\mathrm{CPD}[\mathrm{tip:Cu(110)}]\\
    -\mathrm{CPD}[\mathrm{tip:SrO}]=3.6\mathrm{~eV}
\end{equation*} 
for the SrO termination, and
\begin{equation*}
	\phi[\mathrm{TiO_2}]=\phi[\mathrm{Cu(110)}]-\mathrm{CPD}[\mathrm{tip:Cu(110)}]-\mathrm{CPD}[\mathrm{tip:TiO_2}]=2.6\mathrm{~eV}
\end{equation*} 
for the TiO$_2$ termination for both measurement temperatures. Multiple experiments of this type were performed and the results for either termination coincide with the absolute assignment of $\phi$[SrO] and $\phi$[TiO$_2$] within $\pm$0.2\,eV.

\raggedbottom
\pagebreak

\section{Additional information to measurements in Figure 3} 

\begin{figure}[h!]
	\begin{center}
		\includegraphics[width=1\columnwidth,clip=true]{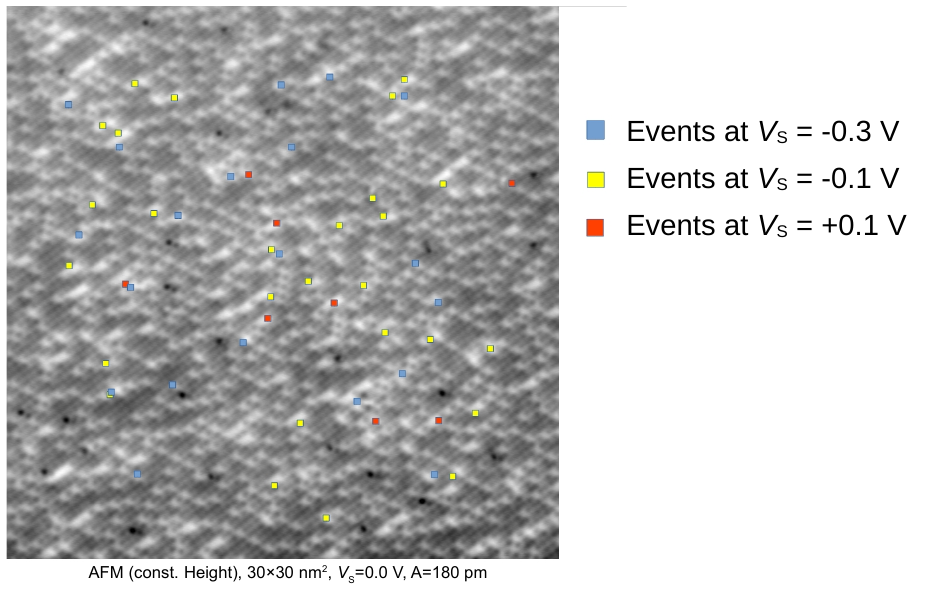}
	\end{center}
	\caption{ \textbf{Locations of hole polarons detected in Figure 3 of the main text.} 
		A constant height AFM image measured at zero bias directly after performing the experiment presented in Figure 3 of the main text. The image was measures at a closer tip-sample distance that allows achieving atomic resolution. The positions of trapped holes were extracted and overlaid on this atomically-resolved image. Squares of different colors mark holes eliminated at the scans with sample biases of -0.3, -0.1 and +0.1 V.  
	}\label{figSM2}
\end{figure}

Figure 3 of the main text showed a procedure for visualizing single holes trapped at surface Sr vacancies. The subtraction of subsequent images was used to highlight the electrostatic forces originating from these holes. All raw data corresponding to this experiment are provided in a supplementary file Figure3RawData.rar. In addition, Figure \ref{figSM2} shows positions of all holes localized in this way. The AFM image in Fig. \ref{figSM2} was measured directly after finishing the discharging experiment presented in Fig.~3 and a closer tip-sample distance was used to achieve good atomic resolution. Here we note that imaging the UV-irradiated surface with such a close sample distance would manipulate or eliminate the photoexcited holes. 

\raggedbottom
\pagebreak

\section{Injection of a measurable tunneling current}

\begin{figure}[h!]
	\begin{center}
		\includegraphics[width=1\columnwidth,clip=true]{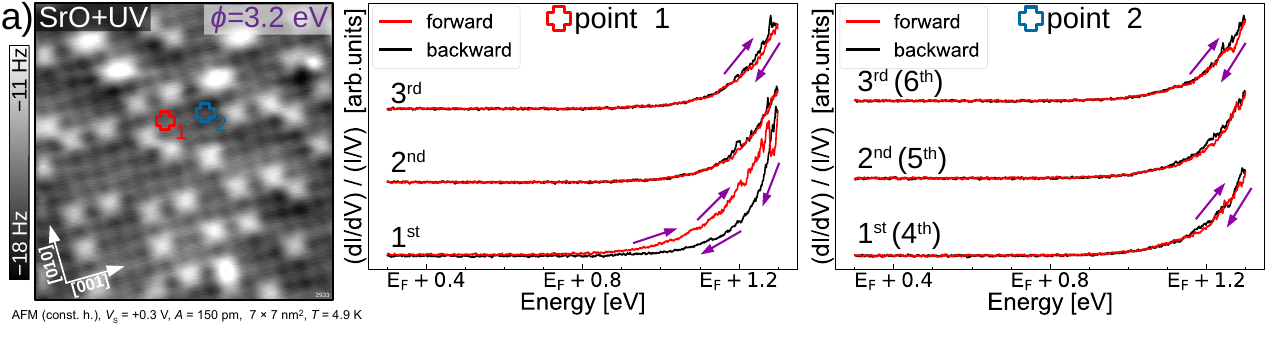}
	\end{center}
	\caption{ \textbf{Hysteresis in the tunneling current} 
		a) AFM iamge of a SrO termination after UV illumination. The tip was positioned to 'Point 1' and a bias sweep was applied, up to the point where a measurable tunneling current appears (few pA). b) STS spectra measured at 'Point 1', where the first spectrum shows hysteresis between the sweeps ramping up and down. Further sweeps at this point do not show any hysteresis. c) The tip was subsequently positioned at 'Point 2', 1 nm apart. The hysteresis is absent.   
	}\label{figSM3}
\end{figure}

In the main text, Figure 3 has shown the single holes trapped at the surface. These can be eliminated by the tip through injection of single electrons, $i.e.$, at conditions where no measurable tunneling current is detected. However, the final discahrging of the surface requires a nonzero tunneling current, as illustrated in Figure \ref{figSM3}. Here the tip was poistioned at point '1' and three STS spectra were measured subsequently. The bias was swept up to +1.3~V where a measurable tunneling current of a few pA could be detected. The spectrum shows a hysteresis between the ramping up and ramping down the bias. This we attribute to elimination of the photoexcited holes induced by the electrons tunneling into the surface. The effect is nonlocal: The tip was subsequently positiond 1 nm apart (Point '2') and no hysteresis could be identified. We speculate that this hysteresis originates from the elimination of photoexcited holes trapped subsurface (within the depletion layer), judging by the nonlocal character of the discharging. 

\raggedbottom
\pagebreak

\section{Measurements at liquid helium temperature}

\begin{figure}[h!]
	\begin{center}
		\includegraphics[width=1\columnwidth,clip=true]{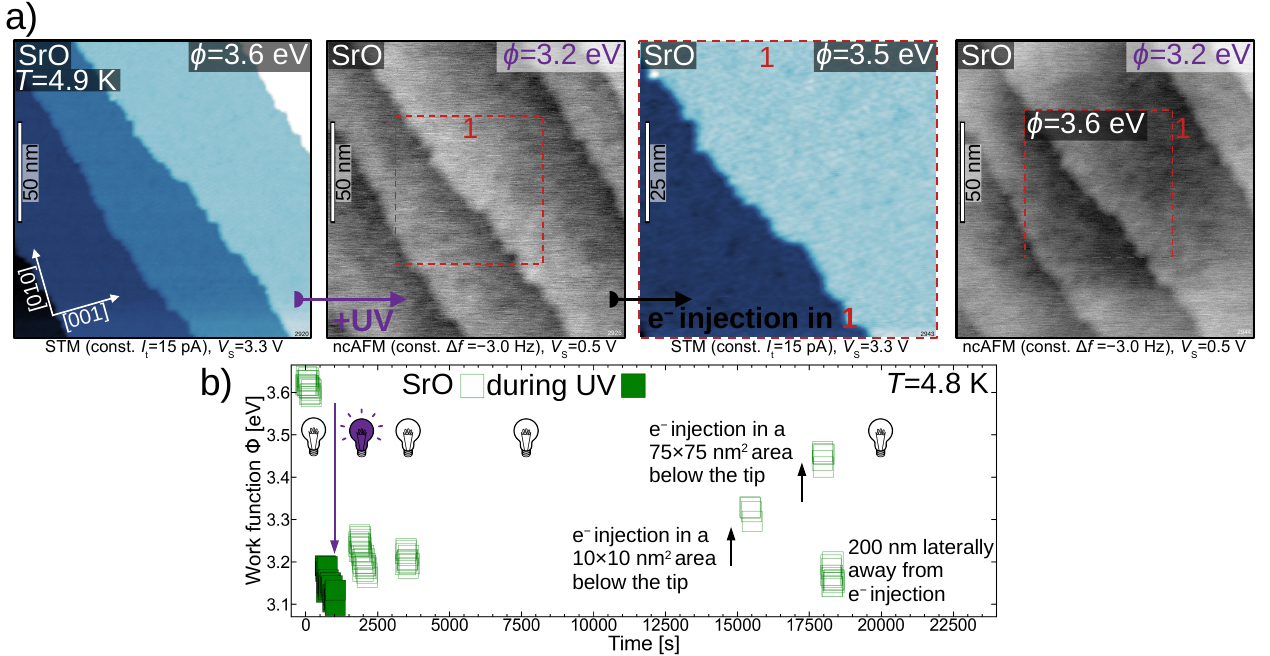}
	\end{center}
	\caption{ \textbf{Measurements at liquid helium} 
		\textit{This experiment is equivalent to the one shown in Figs.\,1,2 of the main text, only performed at LHe temperature.}
        a) Series of feedback-controlled empty-states STM (blue-white color scheme) and feedback-controlled AFM (grayscale) images before and after UV irradiation of the SrO termination at LHe temperature. The rightmost panel demonstrates the spatial localization of the accumulated charge. 
		b) Lifetime of the accumulated h$^+$ following UV irradiation. The work function of the SrO termination decreases upon the UV irradiation, and stays lowered after the UV source is switched off. Only electron injection can return the surface to the ground state.
	}\label{figSM4}
\end{figure}

Figures 1 and 2 of the main text presented data recorded at the temperature of liquid nitrogen ($\approx$78~K). The whole process works similarly at the temperature of liquid helium, see Fig. \ref{figSM4}. The main difference is that the changes of the work function are lower at LHe than at the LN2 temeprature. At LHe, we observed a photovoltage ranging from 0.4 to 0.6 eV, while the LN$_2$ temperature typically provided a photovoltage 0.8 eV. This observation is reproducible over a large number of experiments. We outline two possible reasons: First, the change may arise from the difference in the dielectric permittivity. Upon cooling, the permittivity of SrTiO$_3$ increases, which would lead to a reduced initial band bending and, thus, lower photovoltage. Second, the mobility of the photoexcited charge carriers may be lower at the LHe temperature and they may get trapped easier in the bulk. With the available experimental data, we cannot decide the mechanism with full confidence.

\raggedbottom
\pagebreak

\section{Long lifetime of the photoexcited states}

\begin{figure}[h!]
	\begin{center}
		\includegraphics[width=1\columnwidth,clip=true]{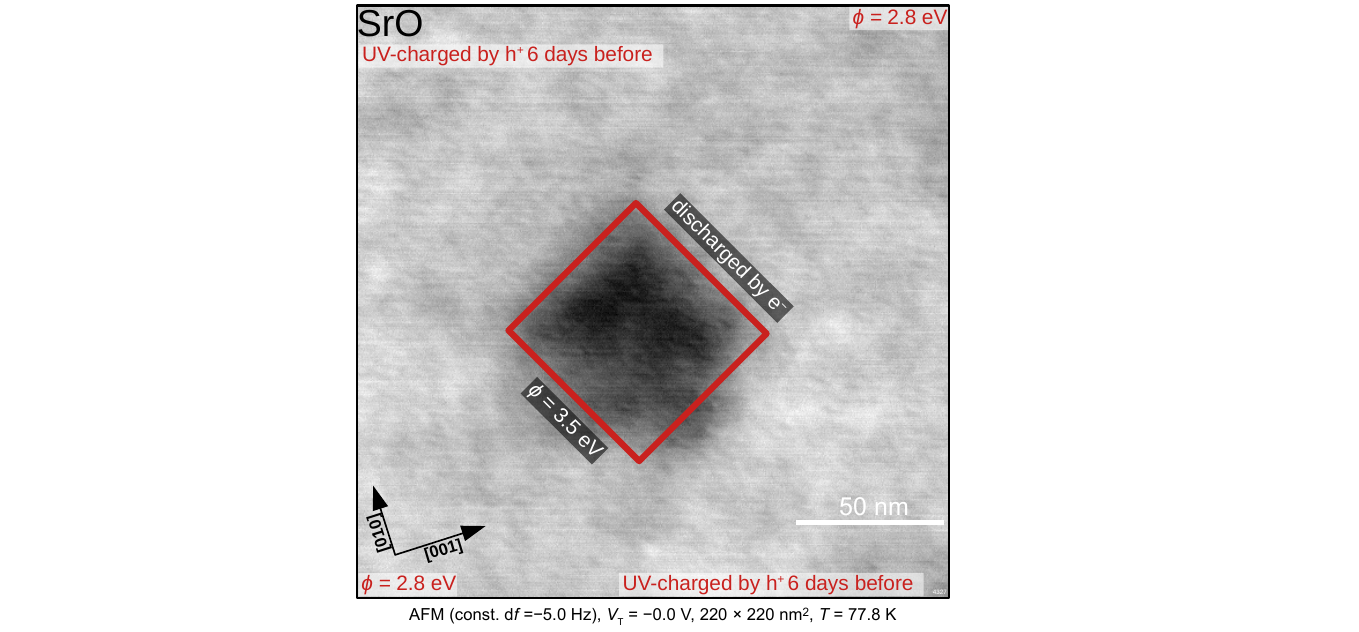}
	\end{center}
	\caption{ \textbf{Accumulated h$^+$ on the UV-irradiated SrO termination are stable for a minimum of 6 days at LN$_2$ temperature.}
		A charged SrO surface was investigated 6 days after the UV irradiation. The discharging through empty-states STM was still possible, and the resulting feedback-controlled AFM image shows a depression in the center where the electron injection took place.
	}
\end{figure}

\raggedbottom
\pagebreak


\end{document}